\documentstyle[epsfig,rotate,preprint,floats,tighten,aps,amssymb]{revtex}

\def\bra#1{{\langle #1{\left| \right.}}}
\def\ket#1{{{\left.\right|} #1\rangle}}
\def\bfgreek#1{ \mbox{\boldmath$#1$}}

\begin{document}
\draft

\preprint{\vbox{\noindent\null \hfill ADP-98-62/T329\\ 
                         \null \hfill hep-lat/9810005 \\
}}

\title{\huge Nucleon Magnetic Moments\\ Beyond the Perturbative Chiral Regime}
\author{Derek B. Leinweber, Ding H. Lu, and Anthony W. Thomas}
\address{Department of Physics and Mathematical Physics\break
 and
        Special Research Centre for the Subatomic Structure of Matter,\break
        University of Adelaide, Australia 5005}
\maketitle

\begin{abstract}
The quark mass dependence of nucleon magnetic moments is explored over
a wide range.  Quark masses currently accessible to lattice QCD, which
lie beyond the regime of chiral perturbation theory ($\chi$PT), are
accessed via the cloudy bag model (CBM). The latter reproduces the
leading nonanalytic behavior of $\chi$PT, while modeling the internal
structure of the hadron under investigation.  We find that the
predictions of the CBM are succinctly described by the simple formula,
$\mu_N(m_\pi) = \mu^{(0)}_N / (1 + \alpha m_\pi + \beta m_\pi^2)$,
which reproduces the lattice data, as well as the leading nonanalytic
behavior of $\chi$PT.  As this form also incorporates the anticipated
Dirac moment behavior in the limit $m_\pi \to \infty$, it constitutes
a powerful method for extrapolating lattice results to the physical
mass regime.
\end{abstract}

\vspace{1.0cm}

\newpage

\section{Introduction}

The fundamental theory of the strong interactions, Quantum
Chromodynamics (QCD), is a nonperturbative field-theory in which the
self-coupling of the gauge field leads to a nontrivial vacuum
structure.  The complexity of the QCD vacuum is manifest in
nonvanishing vacuum expectation values of quark and gluon operator
products having vacuum quantum numbers.  The only known way to
directly calculate the properties of QCD is through the formulation of
a lattice gauge theory where the fields are described on a discrete
space-time lattice.

The lattice formulation of QCD is well established \cite{lattice98}.
Recent advances in lattice action improvement on anisotropic lattices
are greatly facilitating the reduction of systematic uncertainties
associated with the finite lattice volume and the finite lattice
spacing.  However, direct simulation of QCD for light current quark
masses, near the chiral limit, remains computationally intensive.  As
such, the present approach of calculating the properties of QCD using
quark masses away from the chiral regime and extrapolating to the
physical world is likely to persist for the foreseeable future.

The difficulty associated with this approach is illustrated by the
rapid rise of the pseudoscalar mass for small increases in the quark
mass away from the chiral limit as governed by
\begin{equation}
m_\pi^2 = -2 m_q\, {\bra{0} \overline{q} q \ket{0} \over f_\pi^2 },  
\label{GOR} 
\end{equation}
where $q$ denotes a light quark.  Typical quark masses considered in
lattice simulations place $m_\pi \sim 500$ MeV, a value significantly
larger than the physical pion mass, $\mu= 139.6$ MeV.  A mass the
order of 500 MeV is sufficient to largely suppress the pion cloud
contribution to hadronic observables, whereas near the chiral limit,
the pion cloud can make a significant contribution to
them\cite{dblChiCorr}.  Hence it is imperative to extrapolate any
observable simulated on the lattice to the physical world using a
function motivated by the physics of the pion cloud.  Since the pion
cloud contributions are small in lattice simulations, one has to look
beyond the lattice simulation results at present.

Historically, lattice results were often linearly extrapolated with
respect to $m_\pi^2$ to the chiral limit, particularly in exploratory
calculations.  More recently the focus has turned to chiral
perturbation theory ($\chi$PT), which provides predictions for the
leading nonanalytic quark-mass dependence of observables in terms of
phenomenological parameters \cite{golterman94,labrenz94,bernard92,%
sharpe92,holstein98,langacker73,weinberg79,gasser84,gasser88}.

Application of $\chi$PT to the extrapolation of lattice simulation
data is now standard for hadron masses and decay constants
\cite{lattice98}.  However, earlier attempts \cite{butler} to apply
$\chi$PT predictions for the quark-mass dependence of baryon magnetic
moments failed, as the higher order terms of the chiral expansion
quickly dominate the truncated expansion as one moves away from the
chiral limit.  To one meson loop, $\chi$PT expresses the nucleon
magnetic moments as \cite{jenkins93,durand97}
\begin{equation}
\mu_N = \mu_0 + c_1 \, m_\pi + c_2 \, m_\pi^2 \log m_\pi^2 + 
c_3 \, m_\pi^2 + \cdots \, ,
\end{equation}
where $\mu_0$ and $c_3$ are fitted phenomenologically and $c_1$ and
$c_2$ are predicted by $\chi$PT.  The $m_\pi^2 \log m_\pi^2$ term
quickly dominates as $m_\pi$ moves away from the chiral limit making
contact with the lattice results untenable.

Lattice QCD results for baryon magnetic moments
\cite{dblOctet,wilcox92,dong97} remain predominantly based on linear
quark mass (or $m_\pi^2$) extrapolations of the moments expressed in
natural magnetons.  This approach systematically underestimates the
measured moments by 10 to 20\%.  Finite lattice volume and spacing
errors are expected to be some source of systematic error.  However,
$\chi$PT clearly indicates the linear extrapolation of the simulation
results is also suspect.  As such, it is imperative to find a method
which can bridge the void between the realm of $\chi$PT and lattice
simulations.

We report such a method, which provides predictions for the quark mass
(or $m_\pi^2$) dependence of nucleon magnetic moments well beyond the
chiral limit.  In particular, we use the cloudy bag model (CBM), which
involves a relativistic quark model (the MIT bag) coupled to the pion
field in such a way as to restore chiral symmetry
\cite{CBM,thomas84}. The corresponding pion loop corrections to
physical observables reproduce the leading non-analytic behavior of
$\chi$PT.  However, the loop which gives rise to the leading
non-analytic behavior involves two pion propagators.  Because the loop
is regulated by a form factor, related to the finite size of the
hadron, its contribution is suppressed like $1/m_\pi^4$ as $m_\pi$
becomes large. This feature makes it possible to address the larger
quark masses simulated in lattice QCD in a convergent way.

With some tuning of the bag radius, the form factor at the $\pi NN$
vertex and the current quark mass, within the framework of the CBM, we
find that it is possible to reproduce the lattice QCD simulations of
nucleon magnetic moments and smoothly extrapolate to the
experimentally measured values.  We then propose a simple
phenomenological relation designed to reproduce the leading
nonanalytic structure of $\chi$PT and provide the Dirac-moment mass
dependence in the heavy quark-mass regime.  Finally, we illustrate how
such a relation can be used in future lattice QCD calculations.  These
results may also be useful in clarifying issues surrounding the higher
order terms of the chiral expansion of $\chi$PT.

\section{Lattice QCD Simulation Data}

We consider two independent lattice simulations of the nucleon
electromagnetic form factors.  Both calculations employ three-point
function based techniques \cite{loops} utilizing the conserved vector
current such that no renormalization is required in relating the
lattice results to the continuum.  Ref.\ \cite{dblOctet} utilizes
twenty-eight quenched gauge configurations on a $24 \times 12 \times
12 \times 24$ periodic lattice at $\beta=5.9$, corresponding to a
lattice spacing of 0.128(10) fm.  Moments are obtained from the form
factors at 0.16 GeV${}^2$ by assuming equivalent $q^2$ dependencies
for the electric and magnetic form factors.  Ref.\ \cite{wilcox92}
utilizes twelve quenched gauge configurations on a $16^3 \times 24$
periodic lattice at $\beta=6.0$.  The nucleon mass \cite{cabasino91}
corresponds to a lattice spacing of 0.091(3) fm.  Moments are obtained
from dipole fits to the form factors.  Uncertainties are statistical
in origin and are estimated by a single elimination jackknife
\cite{efron79}.  Despite having different lattice volumes and lattice
spacings, the results from the two calculations agree well within
errors.  However, experience suggests that the lattice spacings and
volumes used in these investigations may give rise to scaling
violations and finite size effects the order of 15 to 20\% from the
infinite volume continuum limit.

Ideally, one would like to perform the analysis of chiral corrections,
using results from full QCD, as opposed to quenched QCD.
Unfortunately, such results are not yet available.  Instead we will
utilize these results under the standard approach of correcting the
lattice scale by fixing the lattice spacing using the nucleon mass.
In the absence of any known way to correct for the nonperturbative
effects of quenching, we will assume that these lattice results are a
reasonable representation of full QCD.  Since the quark masses
simulated on the lattice are somewhat heavy, the dominant
contributions from quark loops neglected in the quenched approximation
are largely perturbative and accounted for in the renormalization of
the lattice spacing.  As we shall see, the errors due to quenching are
likely the same order of magnitude as the statistical uncertainties.

\section{The Cloudy Bag Model}

The linearized CBM Lagrangian with the pseudoscalar pion-quark coupling
(to order $1/f_\pi$ ) is given by \cite{CBM,thomas84} 
\begin{eqnarray}
        \protect{\cal L}
   &=& \left[ \overline q (i\gamma^\mu \partial_\mu-m_q)q - B\right]\theta_V
        - {1\over 2}\overline q q \delta_S \nonumber \\
        && + {1\over 2} (\partial_\mu \bfgreek{\pi})^2
        - {1\over 2} m^2_\pi \bfgreek{\pi}^2
        - {i\over 2f_\pi} \overline q \gamma_5 \bfgreek{\tau} \cdot
        \bfgreek{\pi} q \delta_S, \label{LAG} 
\end{eqnarray}
where $B$ is the bag constant, $f_\pi$ is the $\pi$ decay constant,
$\theta_V$ is a step function (unity inside the bag volume and
vanishing outside) and $\delta_S$ is a surface delta function.  In a
lowest order perturbative treatment of the pion field, the quark wave
function is not affected by the pion field and is simply given by the
MIT bag solution \cite{chodos74a}.  Our calculation is carried out in
the Breit frame with the center-of-mass correction for the bag
performed via Peierls-Thouless projection.  The detailed formulas for
calculating nucleon electromagnetic form factors in the CBM are given
in Ref.~\cite{lu98}.

\begin{figure}[t]
\vspace{1.5cm}
\centering{\
\epsfig{file=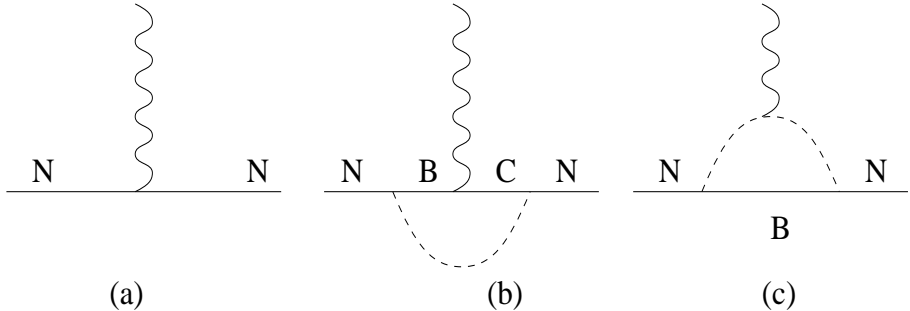,height=4.0cm,width=12cm}
\vspace*{1.0cm}
\caption{Schematic illustration of the processes included in the CBM
calculation.  $B$ and $C$ denote intermediate state baryons and
include $N$ and $\Delta$.}
\label{fig1.ps}}
\end{figure}

In the CBM, a baryon is viewed as a superposition of a bare quark
core and bag plus meson states.  Both the quark core and the meson
cloud contribute to the baryon magnetic moments. These two sources
are balanced around a bag radius, $R = 0.7 - 1.1$ fm \cite{MM}.  
A large bag radius suppresses the contributions from the pion cloud, and
enhances the contribution from the quark core.
The minimal coupling principle is used for the electromagnetic
interaction.  The nucleon magnetic moments can be written as $\mu_N =
G_M^{(q)}(0) + G_M^{(\pi)}(0)$, where $G_M^{(q)}$ is due to $\gamma
qq$ coupling and $G_M^{(\pi)}$ from $\gamma\pi\pi$ coupling.  To one
loop, the CBM reproduces the leading non-analytic 
behavior of $\chi$PT.  The
processes included in this calculation are illustrated in
Fig.~\ref{fig1.ps}.

For the $\pi NN$ vertex, we take the conventional $\pi NN$ coupling
constant, $f_{\pi NN}^2 = 0.0771$.  Instead of the generic form, the
$\pi NN$ form factor is replaced by the phenomenological, monopole
form, $u(k)=(\Lambda^2 - \mu^2)/(\Lambda^2 + k^2)$, where $k$ is the
loop momentum and $\Lambda$ is a cut-off parameter.

In the standard CBM treatment, where the pion is treated as an
elementary field, the current quark mass, $m_q$, is not directly
linked to $m_\pi$.  Most observables are not sensitive to this
parameter, as long as it is in the range of typical current quark
masses.  For our present purpose it is vital to relate the $m_q$
inside the bag with $m_\pi$.  Current lattice simulations indicate
that $m_\pi^2$ is approximately proportional to $m_q$ over a wide
range of quark masses~\cite{cppacs97}.  Hence, in order to model the
lattice results, we scale the mass of the quark confined in the bag as
$m_q= \left(m_\pi/\mu\right)^2 m_q^{(0)}$, with $m_q^{(0)}$ being the
current quark mass corresponding to the physical pion mass.
$m_q^{(0)}$ is treated as an input parameter to be tuned and lies in
the range 6 to 7 MeV.

The parameters of the CBM are obtained as follows.  The bag constant
$B$ and the phenomenological parameter $z_0$ are fixed by the physical
nucleon mass and the stability condition, $dM_N/dR = 0$, for a given
$R_0$ and $m_q^{(0)}$.  For each subsequent value of the pion mass or
the quark mass considered, $\omega_0$ and $R$ are determined
simultaneously from the linear boundary solution of the
bag\cite{chodos74a} and the stability condition, which provides $R^4 =
(3 \omega_0 - z_0)/(4 \pi B)$.  Using the lattice data and the
experimental measurement, the parameters $R_0$, $\Lambda$ and
$m_q^{(0)}$ are tuned to reproduce the experimental moment while
accommodating the lattice data.

\begin{figure}[t]
\centering{\
\epsfig{file=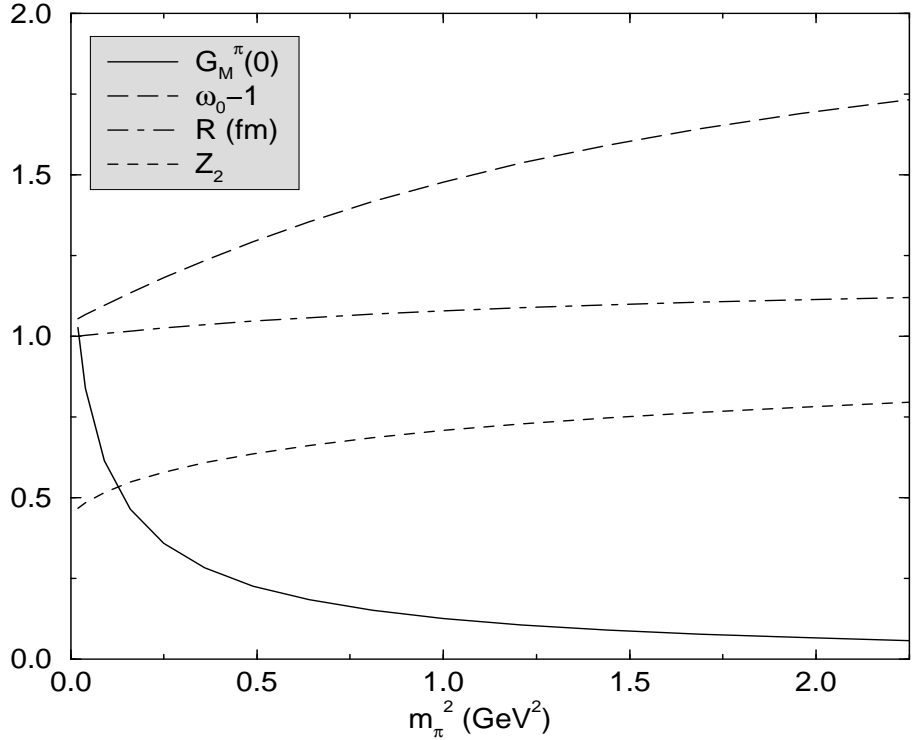,height=10cm,width=12cm}
\caption{The pion contribution to the proton magnetic moment
$G_M^\pi(0)$, and the properties of the bag including the bare bag
probability $Z_2$, the ground state frequency $\omega_0$, and the bag
radius $R$ in variation with the pion mass.}
\label{pion.ps}}
\end{figure}

Fig.~\ref{pion.ps} shows the mass dependence of the bag model
properties and the pion-cloud contribution $G_M^\pi(0)$.  The bare
bag probability, $Z_2$, the quark ground state frequency, $\omega_0$,
and the bag radius, $R$, are plotted as a function of $m_\pi^2$.  As
$m_\pi$ increases, the pion-cloud contribution decreases very quickly
and becomes quite small for large quark masses -- especially in the
range corresponding to the current lattice calculations.  On the other
hand, the bag properties evolve relatively slowly.  As a result, the
dominant influences of $R_0$, $\Lambda$ governing the $\pi NN$ coupling,
and $m_q^{(0)}$ are located in separate regions of $m_\pi$.  The
magnitude of the magnetic moments in the small $m_\pi$ region is
controlled by $\Lambda$ and $R_0$, while a variation of $m_q^{(0)}$ is
more effective in the large $m_\pi$ region, where the pion cloud
nearly vanishes.

\begin{figure}[t]
\centering{\
\epsfig{file=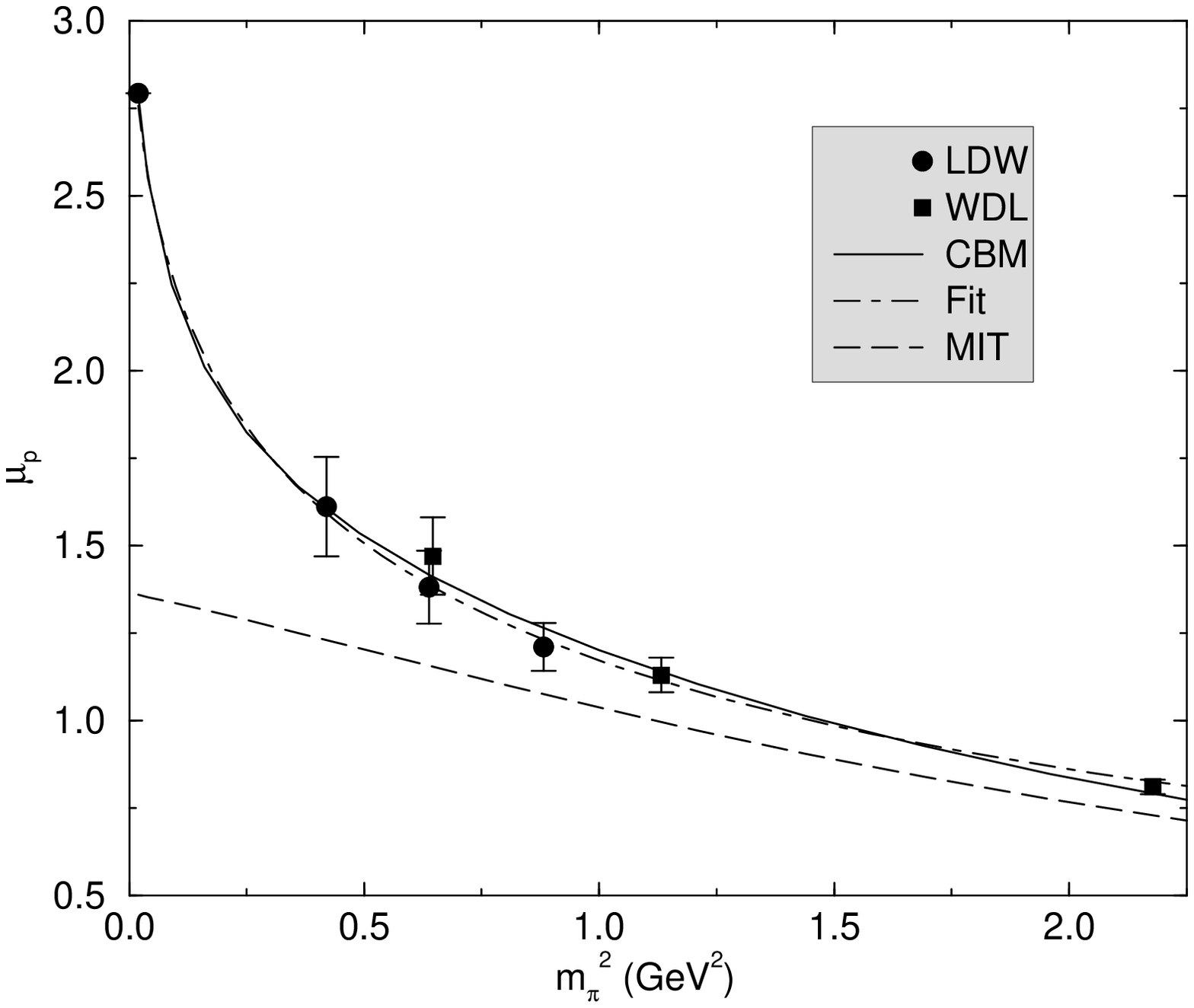,height=10cm,width=12cm}
\caption{The proton magnetic moment as calculated in lattice QCD
($\bullet$ LDW Ref.\ \protect\cite{dblOctet},
$\scriptstyle\blacksquare$ WDL Ref.\ \protect\cite{wilcox92}), the
cloudy bag model (CBM) which includes the pion cloud contribution and
the MIT bag model (MIT) where the pion cloud contribution of the
CBM is omitted.  Also illustrated is a fit of the simple analytic form of
Eq.~(\protect\ref{fit}) to the CBM results.  The point at the physical
value of $m_\pi^2$ is the experimental measurement and is used to
constrain the parameters of the CBM.}
\label{proton.ps}}
\end{figure}

\section{Discussion of Results}

\begin{figure}[t]
\centering{\
\epsfig{file=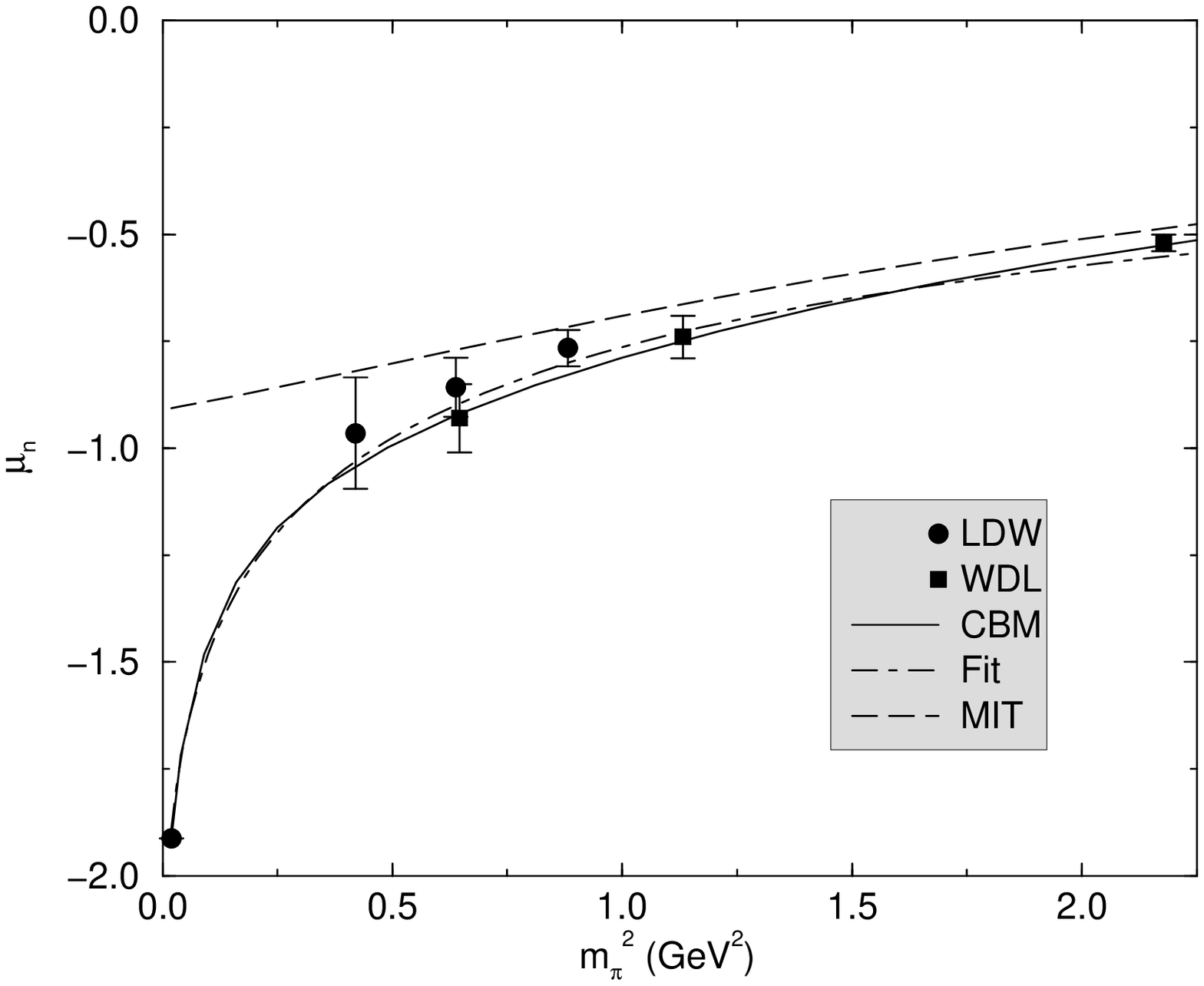,height=10cm,width=12cm}
\caption{The neutron magnetic moment.  Symbols and lines are as
described for Fig.~\protect\ref{proton.ps}.} 
\label{neutron.ps}}
\end{figure}

The nucleon magnetic moments calculated in the CBM are shown in
Fig.~\ref{proton.ps} for the proton, and Fig.~\ref{neutron.ps} for the
neutron by the solid line.  The lattice results ($\bullet$
\cite{dblOctet}, $\scriptstyle\blacksquare$ \cite{wilcox92}) are also
plotted in these figures.  It is possible to simultaneously reproduce
the existing lattice simulation results for the nucleon magnetic
moments at large $m_\pi$ as well as their physical values using CBM
parameters within previously established ranges.  These parameters are
summarized in Table \ref{modelParam}.  The dashed lines in
Figs.~\ref{proton.ps} and \ref{neutron.ps} indicate the corresponding
results for the MIT bag model.  There, without the pion cloud, the
$m_\pi^2$ dependence of the magnetic moments becomes nearly linear.
This clearly shows the significance of the meson cloud, especially in
the small $m_\pi$ regime.

\begin{table}[b]
\caption{CBM parameters and optimal fit parameters for the fit
function of Eq.~(\protect\ref{fit}).  Other CBM parameters are $Z_0 =
2.59$ and $B^{1/4}=144$ MeV. At physical pion mass, the experimental 
magnetic moments (2.79 and -1.91 for the proton and neutron)
are reproduced in the CBM. }
\label{modelParam}
\begin{center}
\begin{tabular}{l|ccc|ccc}
N & $R_0$ (fm) & $\Lambda$ (GeV) & $m_q^{(0)}$ (MeV) & 
 $\mu_N^{(0)}$ & $\alpha$ (GeV$^{-1}$) & $\beta$ (GeV$^{-2}$) \\ 
\hline
proton   &  1.0& 0.68&  4.8 &  3.31 & 1.37 & 0.452 \\
neutron  &  1.0& 0.59&  4.8 & -2.39 & 1.85 & 0.271 \\
\end{tabular}
\end{center}
\end{table}

Since the pion cloud in the quenched approximation is quite different
from that of full QCD
\cite{golterman94,labrenz94,bernard92,sharpe92,dblPiCloud}, one might
regard the difference between the CBM and MIT model results (i.e. the
pion contribution) as indicative of the absolute systematic error
associated with the quenched approximation.  For the quark masses
actually simulated on the lattice, this error is the order of 15\%.
However, some of this difference is already accounted for in the
lattice results through a renormalization of the lattice spacing used
in expressing the moments in nuclear magnetons.  Hence, for the
purposes of this investigation, it is reasonable to accept the
quenched results as an approximate representation of the full QCD
result.

\section{Encapsulating Form}

Having established the quark mass dependence of the nucleon moments
over a very wide range, we now turn to encapsulating these results in
a simple analytic form that might be used in future lattice QCD
extrapolations of simulation results.
The following function
\begin{equation}
  \mu_N(m_\pi) = {\mu_N^{(0)} \over 1 + \alpha m_\pi + \beta m_\pi^2}
  \, ,
\label{fit}
\end{equation}
is matched to the CBM model results by tuning the three parameters
$\mu_N^{(0)}$, $\alpha$ and $\beta$.  These fits are also illustrated
in figures \ref{proton.ps} and \ref{neutron.ps}.  This functional form
provides the correct limiting behavior as a function of $m_\pi$.  As
$m_\pi \to 0$, Eq.~(\ref{fit}) may be expanded as
\begin{equation}
\mu_N(m_\pi) = \mu_N^{(0)} \left [ 1 - \alpha m_\pi + 
(\alpha^2 - \beta) m_\pi^2 + \cdots \right ],
\label{smallmpi}
\end{equation}
such that the leading nonanalytic behavior is proportional to $m_\pi$
as required by $\chi$PT.  For large $m_\pi$, Eq.~(\ref{fit}) leads to
\begin{equation}
\mu_N(m_\pi) = {\mu_N^{(0)}\over \beta m_\pi^2}
\left (1 - {\alpha \over \beta m_\pi} + \cdots \right) \, ,
\end{equation}
such that the magnetic moments decrease as $1/m_q$ for increasing quark
mass, precisely as the Dirac moment requires.  In short, this function
reproduces the leading nonanalytic behavior of $\chi$PT, provides the
desired mass dependence of the Dirac moment in the heavy quark mass
limit and also has the required shape in our region of interest.  The
resulting parameters for this fit are also listed in
Table~\ref{modelParam}.

Because the leading nonanalytic terms predicted in $\chi$PT for baryon
magnetic moments are proportional to the pseudoscalar meson mass, the
kaon cloud is often regarded as the most important and dominant
contribution.  Instead, in the CBM with its natural high momentum
cut-off, the overall contribution of the kaon loop is strongly
suppressed.  Moreover, the kaon cloud contribution to the nucleon
magnetic moments will not display the same significant curvature
associated with the pion cloud.  In consequence, the kaon cloud
contributions may simply be absorbed into the fit parameters
$\mu_N^{(0)}$ and $\beta$.

\begin{table}[t]
\caption{Values for the coefficient of the leading nonanalytic
term obtained from the chiral expansion of Eq.~(\protect\ref{fit})
adjusted to fit the CBM results for the proton and neutron and from
$\chi$PT.  $\mu_N^{(0)}$ and $\alpha$ are from Table
\protect\ref{modelParam}.  $D_0$ and $F_0$ are tree-level coefficients
($D_0 + F_0 = g_A = 1.27$) while $D_1$ and $F_1$ are the one-loop
corrected estimates of Ref.~\protect\cite{jenkins93}. Units are
GeV${}^{-1}$.  In the column headings, the upper and lower signs
correspond to the proton and neutron respectively.}
\label{chiPTcomp}
\begin{center}
\begin{tabular}{lccc}
Nucleon &Encapsulating Form   
        &\multicolumn{2}{c}{Chiral Perturbation Theory} \\
        &$ - \mu_N^{(0)} \alpha$ 
        &$\mp \displaystyle{m_N (D_0 + F_0)^2 \over 8 \pi f_\pi^2}$  
        &$\mp \displaystyle{m_N (D_1 + F_1)^2 \over 8 \pi f_\pi^2}$  \\
\hline
proton   &$-$4.54 &$-$6.97 &$-$4.41 \\
neutron  &   4.42 &   6.97 &   4.41 \\
\end{tabular}
\end{center}
\end{table}

To evaluate the ability of Eq.~(\ref{fit}) to encompass the CBM
predictions, the lattice QCD data and maintain the correct leading
nonanalytic behavior of $\chi$PT, we compare the coefficient of
$m_\pi$ in Eq.~(\ref{smallmpi}) with that predicted by $\chi$PT.  Table
\ref{chiPTcomp} summarizes numerical values for this coefficient.  The
one-loop corrected estimates for the $D$ and $F$ coefficients provide
better agreement between $\chi$PT and experiment for many observables
\cite{jenkins93}.  The similarity between this $\chi$PT estimate and
the coefficient from the encapsulating form is encouraging.

\section{Extrapolation of Lattice QCD Data}

\begin{figure}[t]
\centering{\
\rotate{\epsfig{file=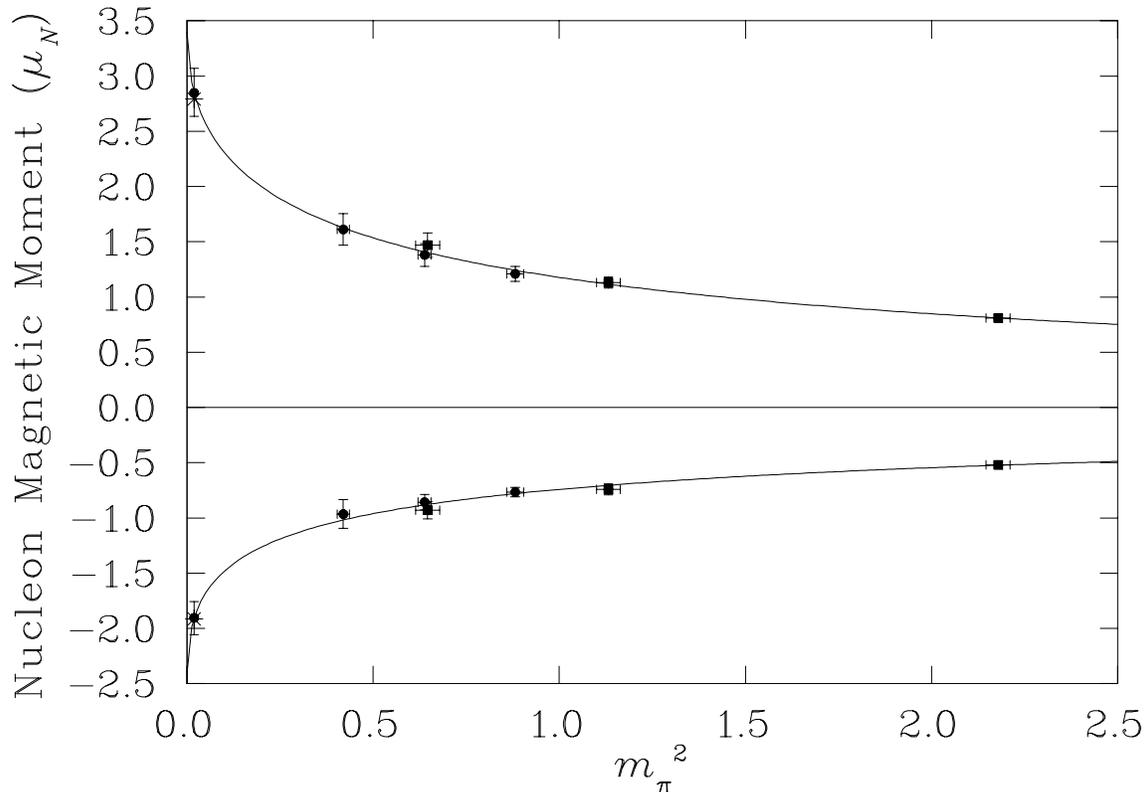,height=15cm}}
\caption{Extrapolation of lattice QCD magnetic moments ($\bullet$
LDW Ref.\ \protect\cite{dblOctet}, $\scriptstyle\blacksquare$ WDL
Ref.\ \protect\cite{wilcox92}) for the proton (upper) and
neutron (lower) to the chiral limit.  The curves illustrate a
two parameter fit of Eq.~(\ref{fit}) to the simulation data in which
the one-loop corrected chiral coefficient of $m_\pi$ is taken from
$\chi$PT.  The experimentally measured moments are indicated by
asterisks.}
\label{NucleonMomFit}}
\end{figure}

Future lattice QCD studies of octet baryon magnetic moments will make
better contact with experiment by adopting the one-loop corrected
coefficient of $m_\pi$ from $\chi$PT and performing a two parameter
fit of Eq.~(\ref{fit}) to the simulation data \cite{comment}.  The
utility of this approach is illustrated in Fig.\ \ref{NucleonMomFit}.
The nucleon magnetic moments at the physical pion mass obtained from this
extrapolation are
\begin{equation}
\mu_p = 2.85(22)\ \mu_N \quad \mbox{and} \quad \mu_n = -1.90(15)\ \mu_N
\end{equation}
and agree surprisingly well with the experimental measurements, 2.793
and $-$1.913 $\mu_N$ respectively.  The fit parameters $(\mu^{(0)},
\beta)$ in units of $(\mu_N, \mbox{GeV}^{-2})$ are $(3.39(23),
0.58(16) )$ and $(-2.40(16), 0.41(16) )$ for the proton and neutron
respectively.  We note that the data required to do a fit of the
lattice results in which covariances are taken into account is no
longer available.  As such, the uncertainties quoted here should be
regarded as indicative only.

\section{Conclusions}

In summary, we have explored the quark mass dependence of nucleon
magnetic moments.  Quark masses beyond the regime of chiral
perturbation theory have been accessed via the cloudy bag model which
reproduces the leading nonanalytic behavior of $\chi$PT and provides
internal structure for the hadron under investigation.  We find that
the predictions of the CBM are succinctly described by a simple
formula which reproduces the leading nonanalytic behavior of $\chi$PT
in the limit $m_\pi \to 0$ and provides the anticipated Dirac moment
behavior in the limit $m_\pi \to \infty$.  The significance of
nonlinear behavior in extrapolating nucleon magnetic moments as a
function of $m_q$ to the chiral regime has been evaluated.  We find
that the leading nonanalytic term of the chiral expansion dominates
from the chiral limit up to the physical pion mass.  Beyond the
physical mass, higher order terms become important and dominate.  This
curvature, neglected in previous linear extrapolations of the lattice
data, can easily account for the departures of earlier lattice
estimates from experimental measurements.  As finite volume and
lattice spacing artifacts are eliminated in future lattice QCD
simulations, it will be interesting to see if the fit parameters
adjust accordingly to maintain and perhaps improve the level of
agreement seen in this investigation.  We advocate the use of the
function in Eq.~(\ref{fit}) in future lattice QCD investigations of
octet baryon magnetic moments.

\acknowledgements

We thank Tony Williams for helpful discussions.  This work is
supported by the Australian Research Council.


\begin{thebibliography}{10}
\bibitem{loops} Magnetic moment contributions from the coupling of the
electromagnetic current directly to sea-quark loops are neglected in
these simulations.  However, the net effect from $u$, $d$, and $s$
sea-quark loops is small due to approximate SU(3)-flavor symmetry
where the net loop contribution vanishes.
\bibitem{comment} 
This dependence on $\chi$PT will eventually be
alleviated as better lattice data becomes available at smaller quark
masses allowing a three parameter fit of Eq.~(\protect\ref{fit}).
\end{thebibliography}

\end{document}